# Large magnetic anisotropy predicted for rare-earth free $Fe_{16-x}Co_xN_2$ alloys

Xin Zhao[*], Cai-Zhuang Wang[§], Yongxin Yao, and Kai-Ming Ho

Ames Laboratory, US DOE and Department of Physics and Astronomy, Iowa State University, Ames, Iowa 50011, USA

[*]E-mail: xzhao@iastate.edu; [§]E-mail: wangcz@ameslab.gov

ABSTRACT

Structures and magnetic properties of $Fe_{16-x}Co_xN_2$ are studied using adaptive genetic algorithm and first-principles calculations. We show that substituting Fe by Co in $Fe_{16}N_2$ with Co/Fe ratio $\leq$ 1 can greatly improve the magnetic anisotropy of the material. The magnetocrystalline anisotropy energy from first-principles calculations reaches 3.18 $MJ/m^3$ (245.6 μeV per metal atom) for $Fe_{12}Co_4N_2$, much larger than that of $Fe_{16}N_2$ and is one of the largest among the reported rare-earth free magnets. From our systematic crystal structure searches, we show that there is a structure transition from tetragonal $Fe_{16}N_2$ to cubic $Co_{16}N_2$ in $Fe_{16-x}Co_xN_2$ as the Co concentration increases, which can be well explained by electron counting analysis. Different magnetic properties between the Fe-rich (x $\leq$ 8) and Co-rich (x > 8) $Fe_{16-x}Co_xN_2$ is closely related to the structural transition.

PACS Numbers: 75.30.Gw, 75.50.Ww, 75.50.Bb, 61.50.-f

Permanent-magnets (PM) are important energy and information storage/conversion materials, and play a crucial role in new energy economies. Currently, the most widely used PM are $Nd_2Fe_{14}B$ and $SmCo_5$, both containing rare-earth (RE) elements. Because of the concerns on limited RE mineral resources and RE supplies, there have been increasing interests in discovering strong RE-free PM materials.[1]

In addition to the microstructures, the performance of PM relies on the intrinsic magnetic properties of their crystal structures, such as saturation magnetization, Curie temperature, and magnetocrystalline anisotropy energy (MAE). Among the RE-free magnets, the metastable tetragonal $\alpha''$-$Fe_{16}N_2$ phase of iron nitrides have attracted considerable experimental and theoretical attentions due to the low cost of Fe and high magnetization in $\alpha''$-$Fe_{16}N_2$ thin films.[2] In experiments, a very large value (~ 3 $\mu_B$) of average Fe magnetic moment in $Fe_{16}N_2$ thin film was reported,[2,3] which is much higher than that of pure bulk Fe (2.2 $\mu_B$). The MAE of the tetragonal $Fe_{16}N_2$ phase has been measured by several experiments,[3-7] but the results are not conclusive, ranging from 0.44 to 2.0 MJ/m$^3$. Because of the promising magnetic properties observed in thin films, efforts to synthesize bulk samples of the tetragonal $\alpha''$-$Fe_{16}N_2$ phase have also been made.[8-10] In the $\alpha''$-$Fe_{16}N_2$ structure, the body-centered cubic (bcc) Fe lattice is expanded into a distorted body-centered tetragonal (bct) lattice due to the presence of N. Close to 500 K, the tetragonal $\alpha''$-$Fe_{16}N_2$ phase decomposes into α-Fe and $Fe_4N$ phases,[11] making the thermal stability one of the major issues for its synthesis and application. It has been shown that adding a small amount of third elements, such as Co, Mn, or Ti, can stabilize the $\alpha''$-$Fe_{16}N_2$ phase in thin film samples.[12,13] Whether such doping approach can stabilize the $\alpha''$-$Fe_{16}N_2$ in the synthesis of bulk samples is still under investigation.[14]

Theoretically, many studies have also been devoted to exploring the origin of the magnetic moment and MAE enhancement in $\alpha''$-$Fe_{16}N_2$.[15-17] Using density functional theory (DFT) and quasiparticle self-consistent GW calculations, Ke *et al.* studied the intrinsic magnetic properties of $Fe_{16}N_2$ and reported a magnetic moment of ~2.5 $\mu_B$ and Curie temperature of ~1300 K.[17] For MAE, they obtained a uniaxial magnetocrystalline anisotropy of 1.03 $MJ/m^3$ by local density approximation (LDA) and 0.65 $MJ/m^3$ by generalized gradient approximation (GGA). Calculations by Ke *et al.* also indicate possible improvement of magnetocrystalline anisotropy by substituting a small amount (less than 15%) of Fe by Co or Ti in the $\alpha''$-$Fe_{16}N_2$ phase.[17] Nevertheless, the calculations were done using the $\alpha''$-$Fe_{16}N_2$ structure and the effect of transition metals (TM) substitution on the crystal structures and phase stabilities has not been investigated.

In this paper, we systematically study the structural evolution and magnetic properties of $Fe_{16-x}Co_xN_2$ with x ranging from 0 to 16. We show that substituting minority of Fe by Co results in better magnetic properties, especially a large value of MAE (~ 3.18 $MJ/m^3$) is achieved at the composition of $Fe_{12}Co_4N_2$ (i.e. x = 4). We demonstrate that there is a tetragonal ($Fe_{16}N_2$) to cubic ($Co_{16}N_2$) structural transition in $Fe_{16-x}Co_xN_2$ as the Co concentration is increased. The changes in the magnetic properties with the Co concentration are strongly correlated with the structural changes.

Our crystal structure searches were performed using adaptive genetic algorithm (AGA)[18,19] with real space cut-and-paste operations for generating offspring structures[20]. No constrains were applied on the structure symmetries during the AGA search. We explored the crystal structures of $Fe_{16-x}Co_xN_2$ (18 atoms per unit cell) with x ranging from 0 to 16. During the AGA searches, auxiliary classical potential based on embedded-atom method[21] was used. The first-principles calculations were carried out using spin-polarized density functional theory (DFT). Generalized-gradient approximation (GGA) in the form of PBE[22] implemented in the VASP code[23] is used. Kinetic energy cutoff was set to be 520 eV. The Monkhorst-Pack's scheme[24] was used for Brillouin zone sampling with a k-point grid resolution of $2\pi$ x 0.05 Å$^{-1}$ during the structure searches. In the final structure refinements, a denser grid of $2\pi$ x 0.03 Å$^{-1}$ was used and the ionic relaxations stop when the force on each atom is smaller than 0.01 eV/Å. Intrinsic magnetic properties, such as magnetic moment and MAE, were also calculated by VASP based on the theoretically optimized structures. All symmetry operations are switched off completely when the spin-orbit coupling is turn on. Meanwhile, a much denser k-point grid ($2\pi$ x 0.016 Å$^{-1}$) is used in the MAE calculations to achieve better k-point convergence.

Based on the structures obtained from our AGA searches, we found that mixing Co with Fe can significantly enhance the magnetic anisotropy as compared to that of $Fe_{16}N_2$ when the Co concentration is smaller than that of Fe. A large value of MAE (~ 3.18 MJ/m$^3$) is found in one of the metastable $Fe_{12}Co_4N_2$ structures, which is the highest among the rare-earth free magnets reported so far. In Fig. 1 (a) – (c), the crystal structures of $Fe_{16}N_2$, $Fe_{12}Co_4N_2$ (the one with largest MAE) and $Co_{16}N_2$ are plotted. The $Fe_{12}Co_4N_2$ structure can be considered as a substituted $Fe_{16}N_2$ structure with Fe at the 4d sites being replaced by Co. For $Fe_{16}N_2$, our calculation gives an MAE of 50.1 μeV/Fe atom (or 0.64 MJ/m$^3$), consistent with Ref. 17 where a value of 52 μeV/Fe atom (or 0.65 MJ/m$^3$) was reported from GGA calculations using the theoretically optimized structure. The $Co_{16}N_2$ structure has nearly zero MAE due to its cubic symmetry.

To demonstrate that these structures are dynamically stable, in Fig. 1(d), the phonon density-of-states of these three structures are plotted. The phonon calculation was performed using a supercell approach by the *Phonopy* code,[25] where supercells with sizes of 324 atoms for $Fe_{16}N_2$ and $Fe_{12}Co_4N_2$ and 288 atoms for $Co_{16}N_2$ were used. The results show that there are no negative frequencies in all three structures, indicating these structures are dynamically stable.

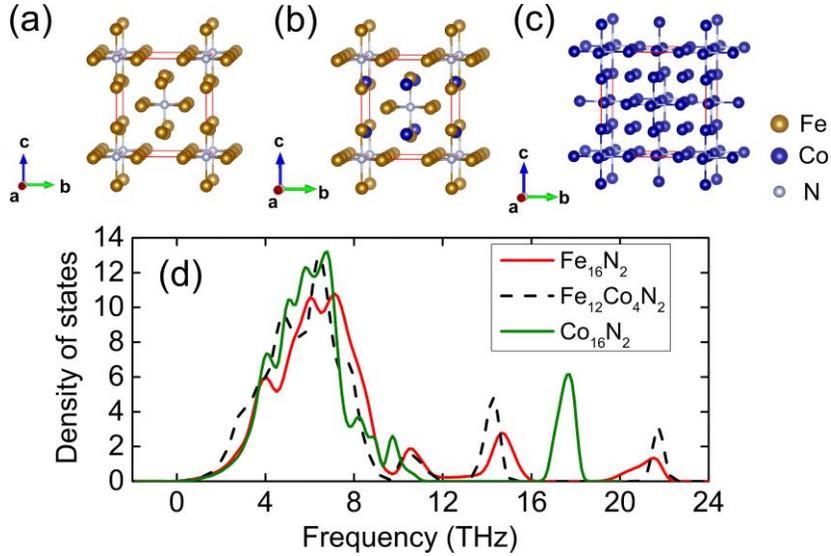

FIG. 1 Crystal structures of (a) $Fe_{16}N_2$, space group *I4/mmm* with $a$=5.68 Å, $c$=6.22 Å and N 2a (0.0, 0.0, 0.0), Fe 4d (0.0, 0.5, 0.25); (b) $Fe_{12}Co_4N_2$, space group *I4/mmm* with a=5.64 Å, c=6.24 Å and N 2b (0.0, 0.0, 0.5), Co 4d (0.0, 0.5, 0.25), Fe 8h (0.255, 0.255, 0.0), Fe 4e (0.0, 0.0, 0.793); (c) $Co_{16}N_2$, space group *Fm-3m* with a=7.19 Å and N 4a (0.0, 0.0, 0.0), Co 24e (0.260, 0.260, 0.260), Co 8c (0.25, 0.25, 0.25). (d) Phonon density-of-states of the structures plotted in (a) – (c).

In addition to the $Fe_{12}Co_4N_2$ structure, other metastable Fe-Co-N structures from our AGA searches also exhibit large MAE. The compositions, energies and MAE of these structures are presented in Fig. 2. Different symbols used in Fig. 2 indicate structures with different MAE values: 1.5 $MJ/m^3$ < MAE < 2.0 $MJ/m^3$ (green squares), 2.0 $MJ/m^3$ < MAE < 3.0 $MJ/m^3$ (black circles) and MAE > 3.0 $MJ/m^3$ (red stars), respectively. From Fig. 2, it is clear that no structure in the Co-rich side has MAE larger than 1.5 $MJ/m^3$. On the contrary, many structures in the Fe-rich side have high MAE. We will show that the trend of MAE is correlated with a structural transition as the function of Co concentration.

Based on the results from our AGA search, the lowest-energy structure of $Co_{16}N_2$ is cubic rather than tetragonal. In Fig. 2, $Co_{16}N_2$ is plotted in a different view so that its relationship to the tetragonal $Fe_{16}N_2$ structure can be seen more clearly. The structure of $Fe_{16}N_2$ has larger distortion than $Co_{16}N_2$, causing the symmetry degraded to a tetragonal space group. In fact, Fe atoms in $Fe_{16}N_2$ have been considered to form distorted bct structure in previous studies.[14,17] In the $Fe_{16}N_2$ structure the bond length between each Fe and its 12 nearest neighbors varies gradually from 2.43 to 3.11 Å. While in the $Co_{16}N_2$ structure, the Co atoms are in a face-centered cubic (fcc) lattice, thus each Co bonds to 12 Co neighbors with almost the same bond length.

In order to build the connection between these two structures, we notice that the fcc structure of $Co_{16}N_2$ ($a$ = 7.19 Å) can also be represented by a bct unit cell with $a' = a/\sqrt{2}$ = 5.08 Å and $c = a$ = 7.19 Å, i.e. the $c/a$ ratio of the tetragonal cell equals to $\sqrt{2}$. The $Fe_{16}N_2$ structure, on the other hand, as being squeezed along $c$ axis, has a larger $a$ (= 5.68 Å) and smaller $c$ (= 6.22 Å). The $c/a$ ratio in the $Fe_{16}N_2$ structure is about 1.09, much smaller than that of $Co_{16}N_2$. Despite the difference in the c/a ratio, these two structures can be classified as the same type of structure, which will be referred to as the $TM_{16}N_2$-type in the following discussions. As shown in Fig. 2, the lowest-energy structures at different compositions all belong to the $TM_{16}N_2$-type structures (represented by the blue squares and connected by the blue line), except the lowest-energy structure of $Fe_8Co_8N_2$ which has a different prototype structure as shown in Fig. 2.

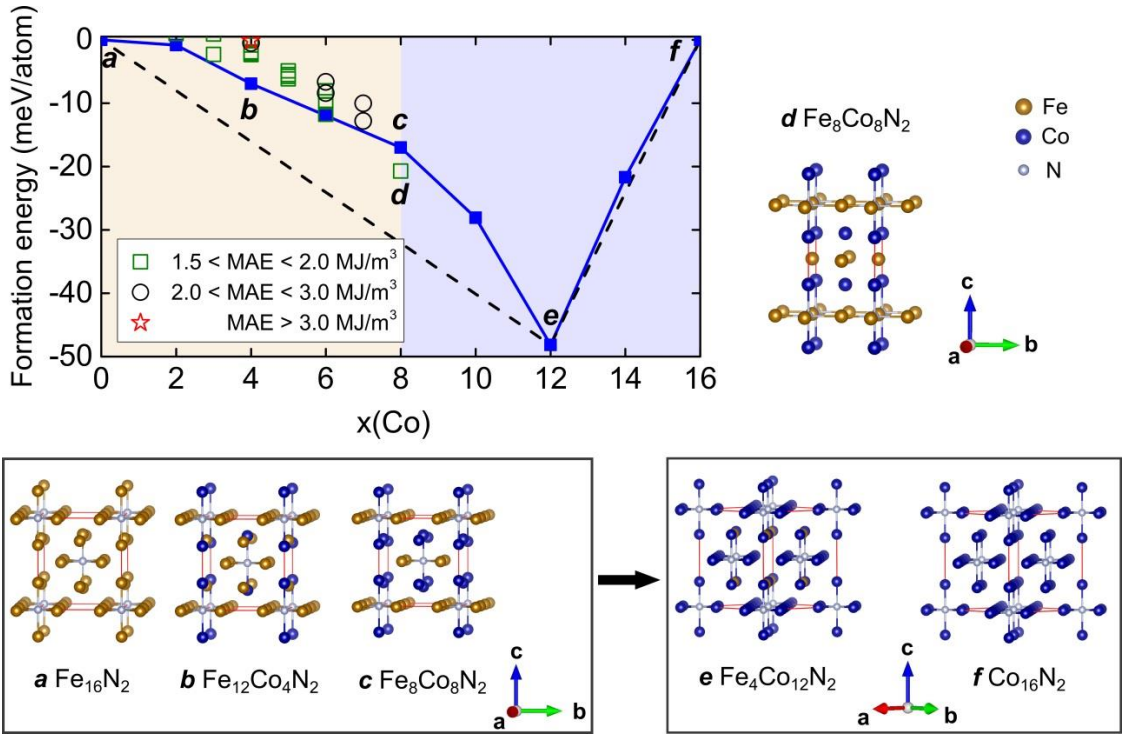

FIG. 2 Formation energies and the structural evolution of $Fe_{16-x}Co_xN_2$ as the function of Co concentration x. The formation energy is calculated using $Fe_{16}N_2$ and $Co_{16}N_2$ as references: $E_f(Fe_{16-x}Co_xN_2) = [E(Fe_{16-x}Co_xN_2) - (1-x/16)*E(Fe_{16}N_2) - x/16*E(Co_{16}N_2)]/18$. The solid blue line connects the $TM_{16}N_2$-type structures (represented by the blue squares, see also text). The convex hull is shown by the black dash line. Calculated structures with high MAE's are indicated by different symbols: green squares ($1.5 < MAE < 2.0$ MJ/m$^3$), black circles ($2.0 < MAE < 3.0$ MJ/m$^3$) and red stars (MAE $> 3.0$ MJ/m$^3$). Crystal structures are plotted for **a**, **b** … **f** as labeled in the formation energy figure.

We note that there is a structural transition as the Co concentration increases. To obtain the transition point, we plot the *c/a* ratio and volume of the $TM_{16}N_2$-type structures (represented by the blue squares in Fig. 2) as a function of Co concentration as shown in Fig. 3(a). For cubic structures such as $Co_{16}N_2$ and $Co_{12}Fe_4N_2$, the *c/a* value is $\sqrt{2}$ as explained above. The results in Fig. 3(a) clearly show a sudden increase in *c/a* at the composition of $Fe_8Co_8N_2$. For the Co-rich side, all the *c/a* ratios are close to $\sqrt{2}$, while on the Fe-rich side including Co:Fe = 1, the c/a drops to around 1.1. Note that the bct structures with *c/a* ratios of 1 and $\sqrt{2}$ are both cubic structures as discussed above. In order to measure the tetragonality of the $Fe_{16-x}Co_xN_2$ structures, we define a parameter $\gamma = \sqrt{|(c/a - 1) \times (\sqrt{2} - c/a)|}$, which is the geometric mean of the distances between the c/a ratio of a given structure from that of the bcc (*c/a*=1) and fcc (*c/a*=$\sqrt{2}$) structures. We found that $\gamma$ is around 0.18 for x ≤ 8, while drops to ~ 0.0 for x > 8, thus serving as a clear indicator of the tetragonal-like to cubic-like structure transition. In addition to c/a, at the transition point from Fe-rich side to Co-rich side, there is nearly 4 percent sudden drop in volume, which can also be consider as the signal of the structure transformation.

The sudden changes in c/a ratio (as well as γ) and volume at the transition can be attributed to the magnetic effects. It is interesting to note that without spin polarization, the above mentioned structure transition disappears as shown by the dashed lines in Fig. 3(a). All structures with different compositions have the c/a ratio close to $\sqrt{2}$ if spin-polarization is turn off in the calculations. At the same time, the volume of the structures is much smaller and increases very slowly as the Co concentration increases. These results indicate that magnetism plays a crucial role for the structural transition in this system.

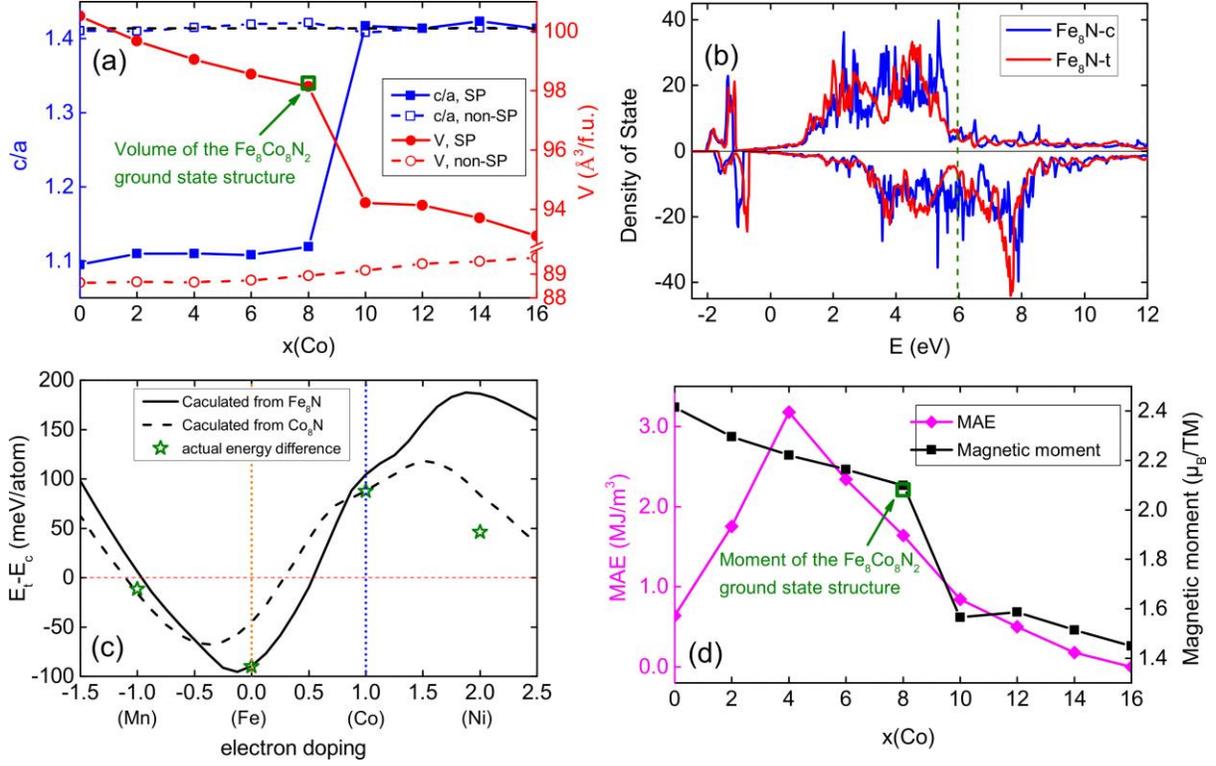

FIG. 3 (a) The *c/a* ratio and volume per formula unit (8 TM atoms, 1 N atom) of the $Fe_{16-x}Co_xN_2$ structures as a function of x from DFT calculations with spin-polarization (solid blue and redlines) and without spin polarization (dash blue and red lines). Black dash line represents $c/a = \sqrt{2}$, which is the value for an ideal fcc structure. (b) Electronic density-of-state of the $Fe_{16}N_2$ in the cubic ($Fe_8N$-c) and tetragonal ($Fe_8N$-t) structures. Fermi level is indicated by the vertical dash lines (two lines are very close). (c) Energy difference between the tetragonal ($E_t$) and cubic ($E_c$) phases evaluated using rigid-band analysis as a function of the electron doping. (d) The largest MAE values obtained at each composition and magnetic moment of the transition metal atoms calculated for the $TM_{16}N_2$-type structures. In plots (a) and (d), the calculated volume and magnetic moment for the lowest energy $Fe_8Co_8N_2$ structure are represented by green open squares.

We further carried out a rigid band perturbation analysis to better understand the mechanism of the structural transition upon TM substitution in $Fe_{16}N_2$. Figure 3(b) shows the electronic density-of-states of $Fe_{16}N_2$ in cubic and tetragonal phases. The electrostatic potential is aligned such that the band energy is equal to the total energy in each system, which was previously used in the development of the tight-binding potentials.[26] By valence electron counting, substitution of one Fe atom with a Mn, Co, or Ni atom is corresponding to doping of 1 hole, 1 electron or 2 electrons, respectively. If all Fe atoms are replaced, the structures will be $Mn_{16}N_2$, $Co_{16}N_2$ and $Ni_{16}N_2$, respectively. The difference of band energy between the tetragonal and cubic phases as the function of electron doping (i.e., the number of valence electron per TM atom relative to that of Fe) in the system is shown by the solid line in Fig. 3(c). The trend predicted by the rigid band analysis is remarkably consistent with the actual total energy calculation results shown by the stars where the atomic relaxations are also included. Similar analyses based on the $Co_{16}N_2$ cubic and tetragonal structures give the same trend, as shown by the dashed line in Fig. 3(c). Therefore the band structure effect through electron or hole doping by substitution of other TM atoms with similar atomic radius in $Fe_{16}N_2$ plays a dominant role in the structural transition of this series of compounds.

The structural transition from tetragonal $Fe_{16}N_2$ to cubic $Co_{16}N_2$ is strongly correlated with the magnetic properties of the system. The magnetic moments of the TM atoms calculated for the lowest-energy $TM_{16}N_2$-type structures are plotted in Fig. 3(d), together with that of the $Fe_8Co_8N_2$ ground-state structure. It can be seen that the behavior of magnetic moment variation is almost the same as that of volume variation. A sudden decrease in magnetic moment near the structural transition point is also observed, which can be attributed to the well-known fact that larger volume gives larger moment.[17,27] For the magnetic anisotropy, as mentioned earlier, after the structure transition, i.e. on the Co-rich side, no structure is found to have MAE larger than 1.5 MJ/m$^3$, while before the transition, structures are found to have much larger MAE's, which can also be seen from Fig. 3(d). This can now be explained from the perspective of crystal structures. The Co-rich structures prefer cubic-like structures, thus unlikely to have strong anisotropy. On the other hand, the shape anisotropy in the Fe-rich tetragonal structures can enhance the MAE.

It should be noted that the magnetic properties of the TM$_{16}$N$_2$-type structures presented in Fig. 3(d) are for stoichiometric compounds. Since the energies of Fe-Co-N alloys with different Fe/Co site occupations are very close to each other, it is likely that at finite temperature disordered site occupation between Fe and Co would occur, thus alloy of Fe$_{16-x}$Co$_x$N$_2$ will form instead of stoichiometric compounds. The effects of Fe and Co occupation disorder on the magnetic properties should be considered in material design and processing. As discussed above, there is a sharp structural phase transition at x=8 between Fe-rich tetragonal and Co-rich cubic structures. Therefore, it is a good approximation to adopt the lattice parameters of tetragonal Fe$_{16}$N$_2$ for the Fe-rich (x≤8) and those of cubic Co$_{16}$N$_2$ for the Co-rich (x≥8) Fe$_{16-x}$Co$_x$N$_2$ alloys. We enumerated all the possible Fe/Co occupations in the tetragonal-Fe$_{16}$N$_2$/cubic-Co$_{16}$N$_2$ structures using a unit cell size of 18 atoms. The atomic positions in the unit cells are fully relaxed by DFT calculations while the lattice parameters of the structures are kept fixed. The results are shown in Fig. 4. As one can see from Fig. 4(a), the energy spread due to the site occupation disorder can be as large as 50-60 meV/atoms. However, the magnetic moment as the function of composition averaged over all the structures at each composition as shown in Fig. 4(b) is very similar to that of the lowest-energy structure plotted in Fig. 3(d). Since the calculation of MAE is very costly, in Fig. 4(b), we plotted the MAE results averaged over 10 lowest-energy configurations for different Co concentrations. Although the averaged MAE's are slightly smaller than the results calculated for stoichiometric compounds, the maximum value appears at the same composition i.e. $Fe_{12}Co_4N_2$. Between the MAE and magnetic moment, it is easy to notice that the error bar in MAE is much larger, indicating that magnetic anisotropy is more sensitive to the change of structures. We have also performed calculations in which both atomic position and lattice parameters are allowed to relax. The behavior of the magnetic moment and MAE as the function of Co concentration is essentially the same as Fig. 4(b) except the transition at x(Co)=8 is smoothed out when the lattice parameters are allowed to relax.

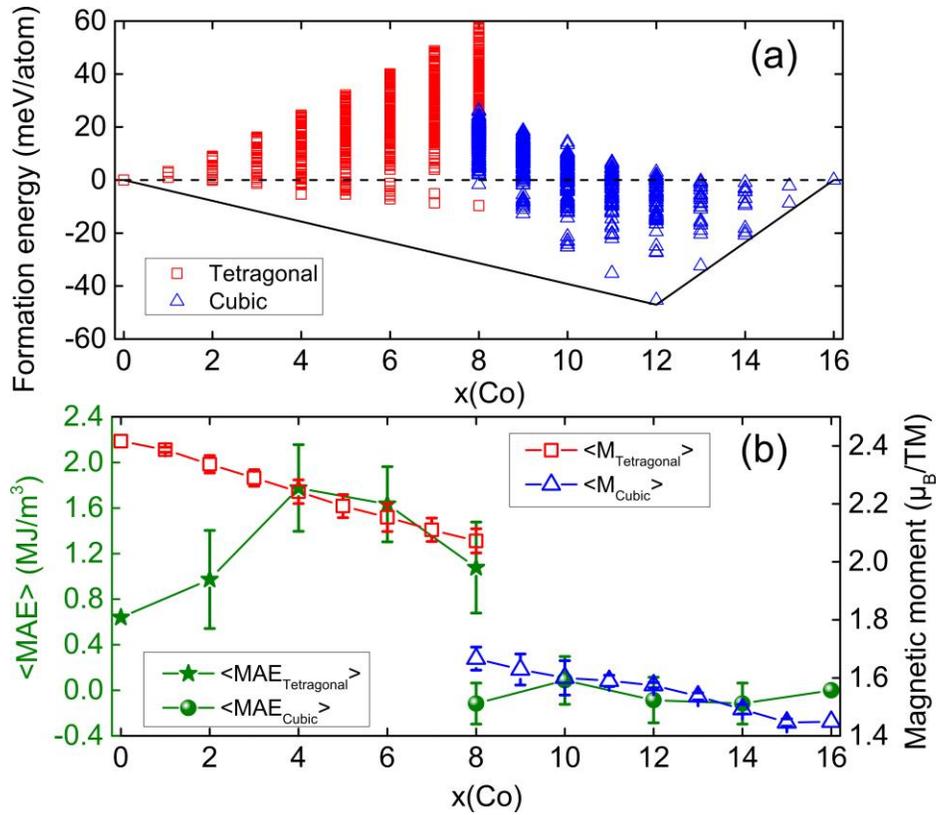

FIG. 4 (a) Formation energies and (b) magnetic properties of the structures obtained from enumerating all the possible Fe/Co occupations in the tetragonal-$Fe_{16}N_2$ (for x(Co) ≤ 8) and cubic-$Co_{16}N_2$ (for x(Co) ≥ 8) structures using a unit cell size of 18 atoms. The energies of tetragonal-$Fe_{16}N_2$ and cubic-$Co_{16}N_2$ structures were used as reference in the formation energy calculation. "< >" in (b) represents the averaged values: the magnetic moments were averaged over all the calculated structures, while the MAE numbers were averaged over 10 lowest-energy structures at each composition. Error bar represents one standard deviation.

In conclusion, our study on $Fe_{16-x}Co_xN_2$ reveals a structure transition as the Co concentration increases, which can be well explained by rigid band perturbation analysis. From the cubic $Co_{16}N_2$ structure to the tetragonal $Fe_{16}N_2$ structure, magnetic moments of the system increase while magnetocrystalline anisotropy is maximized at the composition near $Fe_{12}Co_4N_2$. The substantial improvement in the magnetic properties predicted for both ordered stoichiometric compounds and alloys with disordered Fe/Co occupations makes it possible for $Fe_{16-x}Co_xN_2$ to find applications as a rare-earth free magnet. The stability of this system is still an important issue to consider. Although the structures discussed here are proven to be dynamically stable, we see that using $Fe_{16}N_2$ and $Co_{16}N_2$ (which are metastable already) as references, only $Fe_4Co_{12}N_2$ is stable against decomposition. But it is also noticed that energies of the other structures are within 20 meV/atom above the convex hull, which may be attainable by synthesis techniques at far-from-equilibrium conditions.

ACKNOWLEDGMENT

This work was supported by the National Science Foundation (NSF), Division of Materials Research (DMR) under Award DMREF: SusChEM 1436386. The development of adaptive genetic algorithm (AGA) and the method for rigid band perturbation analysis was supported by the US Department of Energy, Basic Energy Sciences, Division of Materials Science and Engineering, under Contract No. DE-AC02-07CH11358, including a grant of computer time at the National Energy Research Scientific Computing Center (NERSC) in Berkeley, CA.